\documentclass[doublecol]{epl2}

\title{Improved half-metallic ferromagnetism of transition-metal pnictides and chalcogenides
calculated with a modified Becke-Johnson exchange potential}
 \shorttitle{Improved half-metallic ferromagnetism of transition-metal pnictides and chalcogenides}

\author{San-Dong Guo \and Bang-Gui Liu\thanks{E-mail: \email{bgliu@mail.iphy.ac.cn}}}
\shortauthor{San-Dong Guo and Bang-Gui Liu}

\institute{ Institute of Physics, Chinese Academy of
Sciences, Beijing 100190, China\\
Beijing National Laboratory for Condensed Matter Physics, Beijing
100190, China}


\abstract{We use a density-functional-theory (DFT) approach with a
modified Becke-Johnson exchange plus local density approximation
(LDA) correlation potential (mBJLDA) [semi-local,
orbital-independent, producing accurate semiconductor gaps. see F.
Tran and P. Blaha, Phys. Rev. Lett. \textbf{102}, 226401 (2009)]
to investigate the electronic structures of zincblende
transition-metal (TM) pnictides and chalcogenides akin to
semiconductors. Our results show that this potential does not
yield visible changes in wide TM d-$t_{2g}$ bands near the Fermi
level, but makes the occupied minority-spin p-bands lower by
0.25$\sim$0.35 eV and the empty (or nearly empty) minority-spin
$e_g$ bands across the Fermi level higher by 0.33$\sim$0.73 eV.
Consequently, mBJLDA, having no atom-dependent parameters, makes
zincblende MnAs become a truly half-metallic (HM) ferromagnet with
a HM gap (the key parameter) 0.318eV, being consistent with
experiment. For zincblende MnSb, CrAs, CrSb, CrSe, or CrTe, the HM
gap is enhanced by 19$\sim$56\% compared to LDA and generalized
gradient approximation results. The improved HM ferromagnetism can
be understood in terms of the mBJLDA-enhanced spin exchange
splitting.}

\begin{document}

\maketitle

\section{Introduction}

It is believed that half-metallic ferromagnetic materials can be
used to make high-performance spintronic
devices\cite{spintr1,spintr2}, because one of their two spin
channels is metallic and the other has the feature of a
semiconductor or insulator, and then the spin polarization at the
Fermi energy can reach 100\% if the spin-orbit coupling effect is
neglected\cite{hm}. Half-metallic ferromagnetism has been found in
NiMnSb \cite{hm}, CrO$_2$ \cite{cro2}, Fe$_3$O$_4$ \cite{fe3o4},
Co$_2$MnSi \cite{co2mnsi}, and La$_{0.67}$Sr$_{0.33}$MnO$_3$
\cite{lamno3}, and nearly 100\% high spin-polarization has been
observed experimentally in the cases of CrO$_2$ and
La$_{0.67}$Sr$_{0.33}$MnO$_3$ materials\cite{hmsp}. It is of much
interest to seek half-metallic ferromagnetic materials compatible
with important semiconductors. Great efforts have been made in
this direction, and significant progresses have been achieved in
binary transition-metal pnictides and chalcogenides such as MnAs,
MnSb, CrAs, CrSb, CrSe, and CrTe with zincblende
\cite{nhm,mnas,mnas1,mnas2,cras,crsb,crassurf,lbg1,lbg2,zbtm56,pickett6,lbg3,sanyal,crte,mnasab,crte1,mnx}
and wurtzite structures \cite{lbg4}, special
transition-metal-doped semiconductors\cite{sidelta,lgq}, other
transition-metal compounds based on convectional
semiconductors\cite{gete,slj,gsd} etc.

On the theoretical side, one usually uses
density-functional-theory (DFT)\cite{dft} method with generalized
gradient approximation (GGA)\cite{pbe96}, local density
approximation (LDA)\cite{pw92} etc to study the electronic
structures of these materials akin to semiconductors. Because it
is well-known that GGA and LDA underestimate semiconductor gaps,
there naturally exists a question: How good do GGA and LDA
describe the electronic structures of such half-metallic
ferromagnetic materials? This needs to be answered because there
exists some inconsistence between theoretical predictions and
corresponding experimental observations. For examples, zincblende
MnAs has been experimentally shown to be
half-metallic\cite{mnas,mnas1,mnas2}, but both GGA and LDA
calculations prove that it is not truly half-metallic\cite{nhm}
because the Fermi level touches or even crosses the bottom of the
minority-spin conduction bands. Fortunately, there have already
been many feasible approaches to improve DFT calculations of
semiconductor gaps, such as various GW methods\cite{gw,gw1},
exact-exchange approach\cite{exx,exx1,exx2}, different
hybrid-functionals\cite{hybrid}, and a modified Becke-Johnson
(mBJ) exchange potential \cite{mbj09}. Some of them demand high
expenses in computational resources, and some are very economical.
While being as computationally economical as GGA and LDA and
needing no atom-dependent parameters, the mBJ exchange plus LDA
correlation (mBJLDA) potential as an orbital-independent,
semi-local exchange-correlation potential has been proved to
produce accurate gaps for wide-band-gap insulators, sp
semiconductors, and 3d transition-metal oxides\cite{mbj09}.
Therefore, it is highly desirable, even necessary, to use this
greatly-improved, atom-parameters-independent, still economical
DFT approach to investigate such half-metallic materials akin to
semiconductors or insulators.

In this article we use a state-of-the-art DFT method with mBJLDA
potential\cite{mbj09} to investigate the electronic structures of
half-metallic ferromagnets akin to semiconductors. Our results
show that in the cases of zincblende MnAs, MnSb, CrAs, CrSb, CrSe,
and CrTe, mBJLDA does not yield visible changes in wide
transition-metal d-$t_{2g}$ dominating bands near the Fermi level,
but makes the occupied minority-spin p-dominating bands lower by
0.25$\sim$0.35 eV and the empty (or nearly empty) minority-spin
$e_g$ bands higher by 0.33$\sim$0.73 eV compared to GGA and LDA
results. Consequently, mBJLDA makes zincblende MnAs become a truly
half-metallic ferromagnet, being consistent with experiment, and
enhances the half-metallic gaps\cite{lbgb} of the other five by
19$\sim$56\%. The improved half-metallic ferromagnetism can be
understood in terms of the fact that the spin exchange splitting
is enhanced by using mBJLDA potential. This consistence is a
direct evidence that mBJLDA as a semi-local and
orbital-independent exchange-correlation potential without
atom-dependent parameters is very satisfactory in describing the
electronic structures of such materials akin to semiconductors.
More detailed results will be presented in the following.

The rest of the paper is organized as follows. We shall describe
our computational detail in next section. In the third section we
shall present our main results calculated with mBJLDA in
comparison with those calculated with the GGA and LDA. In the
fourth section we shall make necessary discussions and try to
understand the calculated results. Finally, we shall give our
conclusion in the fifth section.

\section{Computational detail}

We use a full-potential linearized augmented-plane-waves method
within the density functional theory \cite{dft}, as implemented in
package WIEN2k \cite{wien2k}. We use mBJLDA for the
exchange-correlation potential\cite{mbj09,pw92} to do our main DFT
calculations, and take popular LDA by Perdew and Wang (LDA-PW91)
\cite{pw92} and GGA by Perdew, Burke, and Ernzerhof
(GGA-PBE)\cite{pbe96} to do corresponding comparative studies. The
mBJ exchange potential\cite{mbj09} was developed from a semi-local
exchange potential proposed by Becke and Johnson (BJ)\cite{bj}.
The BJ exchange potential can reproduce the shape of the
exact-exchange optimized-effective potential of atoms\cite{bj}.
mBJLDA as a combination of mBJ exchange and LDA correlation can
produce accurate semiconductor gaps for sp semiconductors,
wide-band-gap semiconductors, and transition-metal oxide
semiconductors and insulators\cite{mbj09}. The full relativistic
effects are calculated with the Dirac equations for core states,
and the scalar relativistic approximation is used for valence
states \cite{relsa,relsa1,relsa2}. The spin-orbit coupling is
neglected because it has little effect on our results. We use 3000
k points in the first Brillouin zone, make harmonic expansion up
to $l_{\rm max}=10$ in each of the atomic spheres, and set
$R_{mt}*k_{\rm max}=8$. The radii of the atomic spheres of Mn, Cr,
As, Sb, Se, and Te are set to 2.41, 2.41, 2.10, 2.27, 2.16, and
2.33 bohr, respectively. The volumes are optimized by using
LDA-PW91 and GGA-PBE in terms of total energy method. The
self-consistent calculations are considered to be converged when
the integration of absolute charge-density difference between the
input and output electron density is less than 0.0001$|e|$ per
formula unit, where $e$ is the electron charge.

\section{Main calculated results}

At first, we investigate zincblende MnAs to show advantage of
mBJLDA. This is because on the experimental side it has been shown
to be half-metallic and intensively studied for device
applications\cite{mnas,mnas1,mnas2}, but on the theoretical side
both LDA and GGA calculations show that it is not truly
half-metallic\cite{nhm}. Since there is no corresponding energy
functional for the mBJ exchange potential\cite{mbj09}, we use
experimental lattice constant as our computational input. We study
its electronic structures by using the three exchange-correlation
potentials: LDA-PW91, GGA-PBE, and mBJLDA. We present its total
and partial DOSs calculated with mBJLDA, GGA-PBE, and LDA-PW91 in
Fig. 1. The mBJ result shows that zincblende MnAs is a
half-metallic ferromagnet. The total magnetic moment per formula
unit is 4$\mu_B$, in contrast with 3.86$\mu_B$ calculated with
GGA-PBE. The minority-spin gap across the Fermi level is 2.352 eV,
substantially larger than the GGA-PBE result, 1.76 eV, and the
half-metallic gap is determined by the bottom of the minority-spin
conduction bands, being 0.318 eV. Similar results could be
obtained by adding appropriate orbital-dependent on-site
correlation parameters to some localized orbitals, as was done in
LDA+U approach \cite{ldau} in the case of
Co$_2$FeSi\cite{co2fesi}. Another special feature in the mBJLDA
DOS is that the As-p-dominating bands are merged with the
Mn-d-dominating bands in the majority-spin channel, and the
As-p-dominating bands in both spin channels are a little narrower
than those calculated with both GGA-PBE and LDA-PW91. The mBJLDA
DOS result implies 100\% spin polarization at the Fermi level, but
either LDA-PW91 or GGA-PBE result leads to much smaller spin
polarization at the Fermi level. Therefore, the mBJLDA DOS is
consistent with experiment, but either of the LDA-PW91 and GGA-PBE
results is not.

\begin{figure}[!htb]
\begin{center}
\includegraphics[width=8cm]{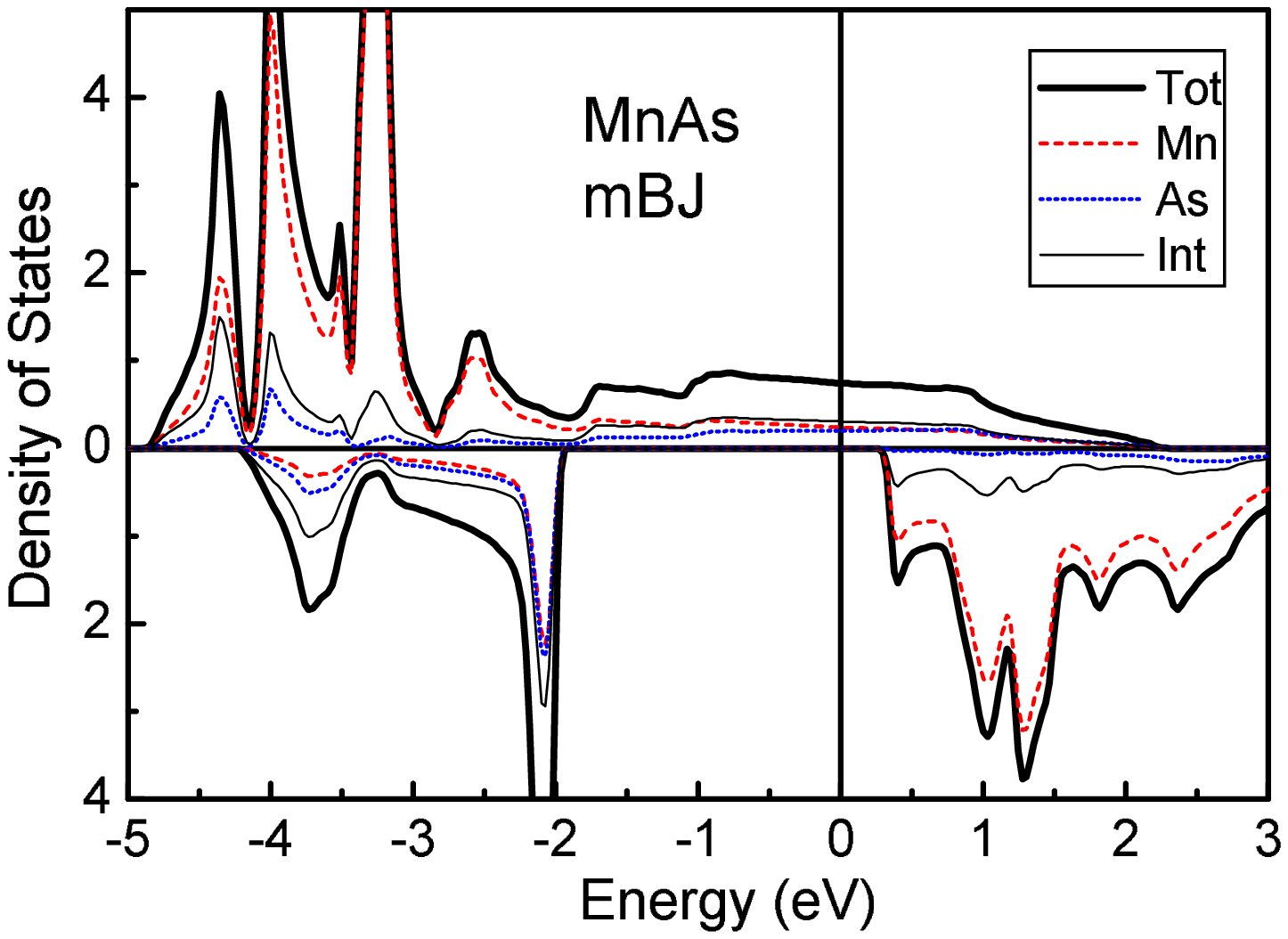}\\
\includegraphics[width=8cm]{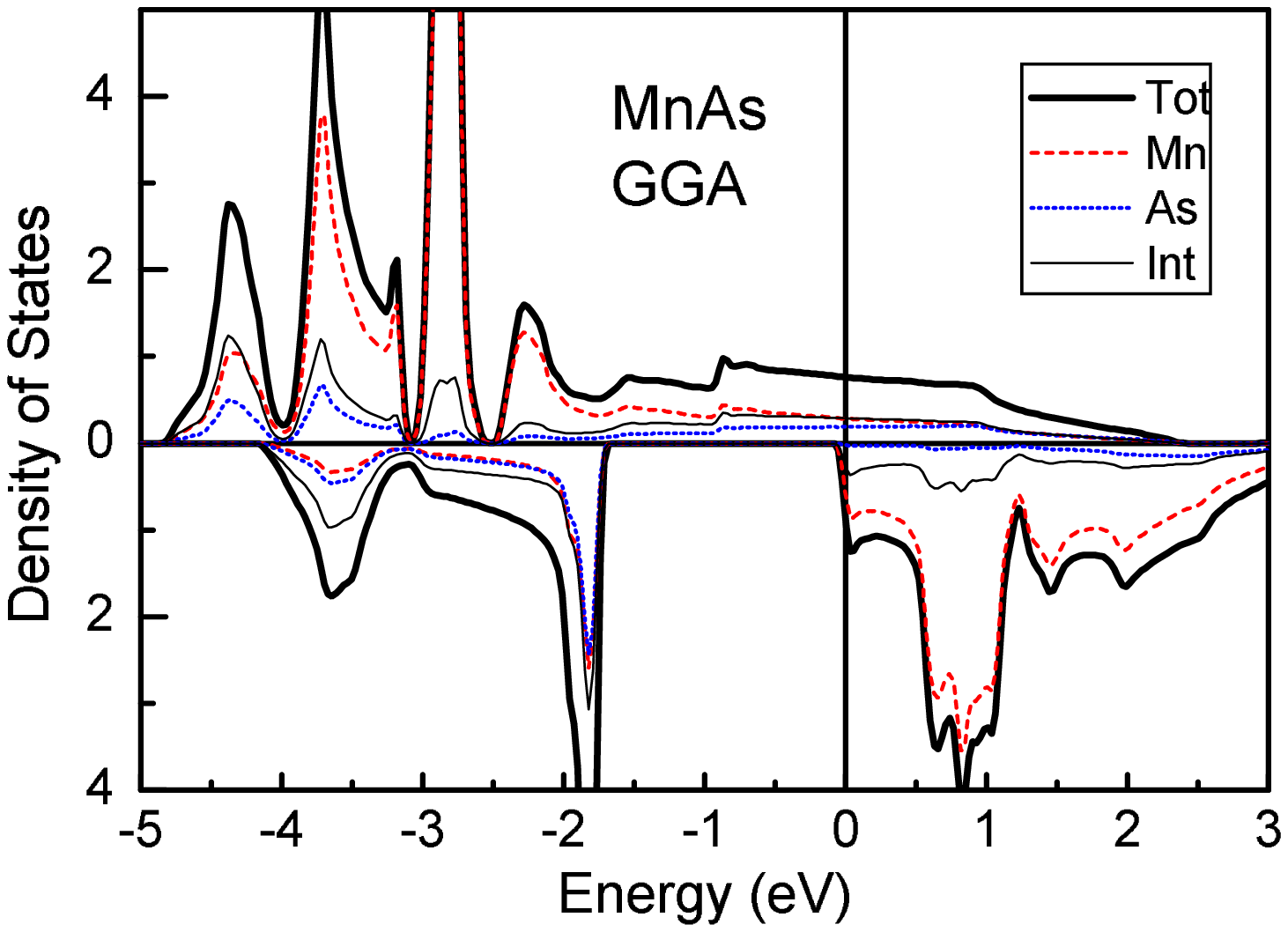}\\
\includegraphics[width=8cm]{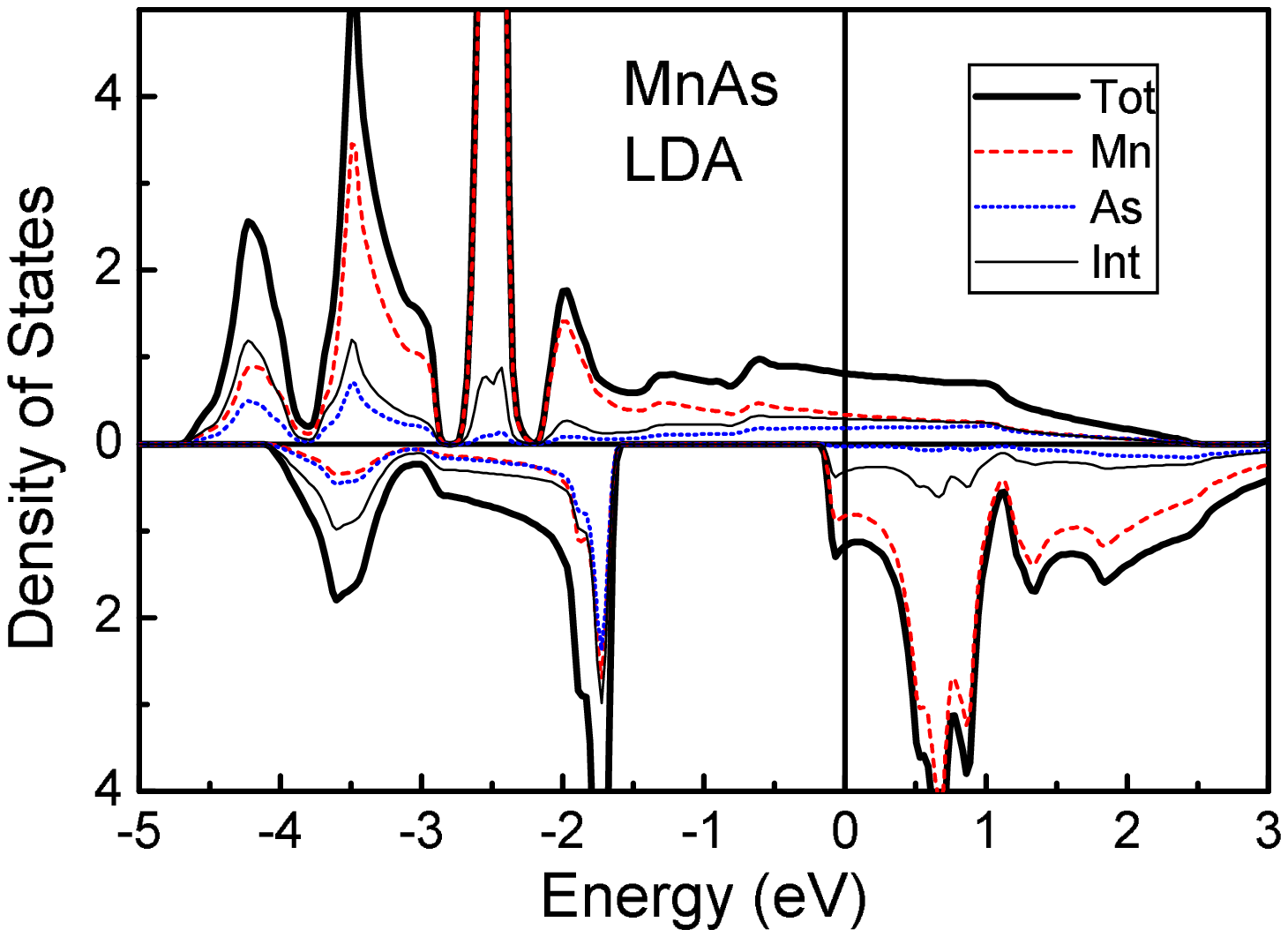}
\caption{(color online) Spin-resolved density of states of MnAs
with the zincblende structure, calculated with mBJLDA (upper
panel), GGA-PBE (middle panel), and LDA-PW91 (lower panel),
respectively. The upper part in each panel is for majority-spin
channel and the lower part for minority-spin.}\label{mnas-dos}
\end{center}
\end{figure}

\begin{figure}[!htb]
\begin{center}
\includegraphics[width=4.2cm]{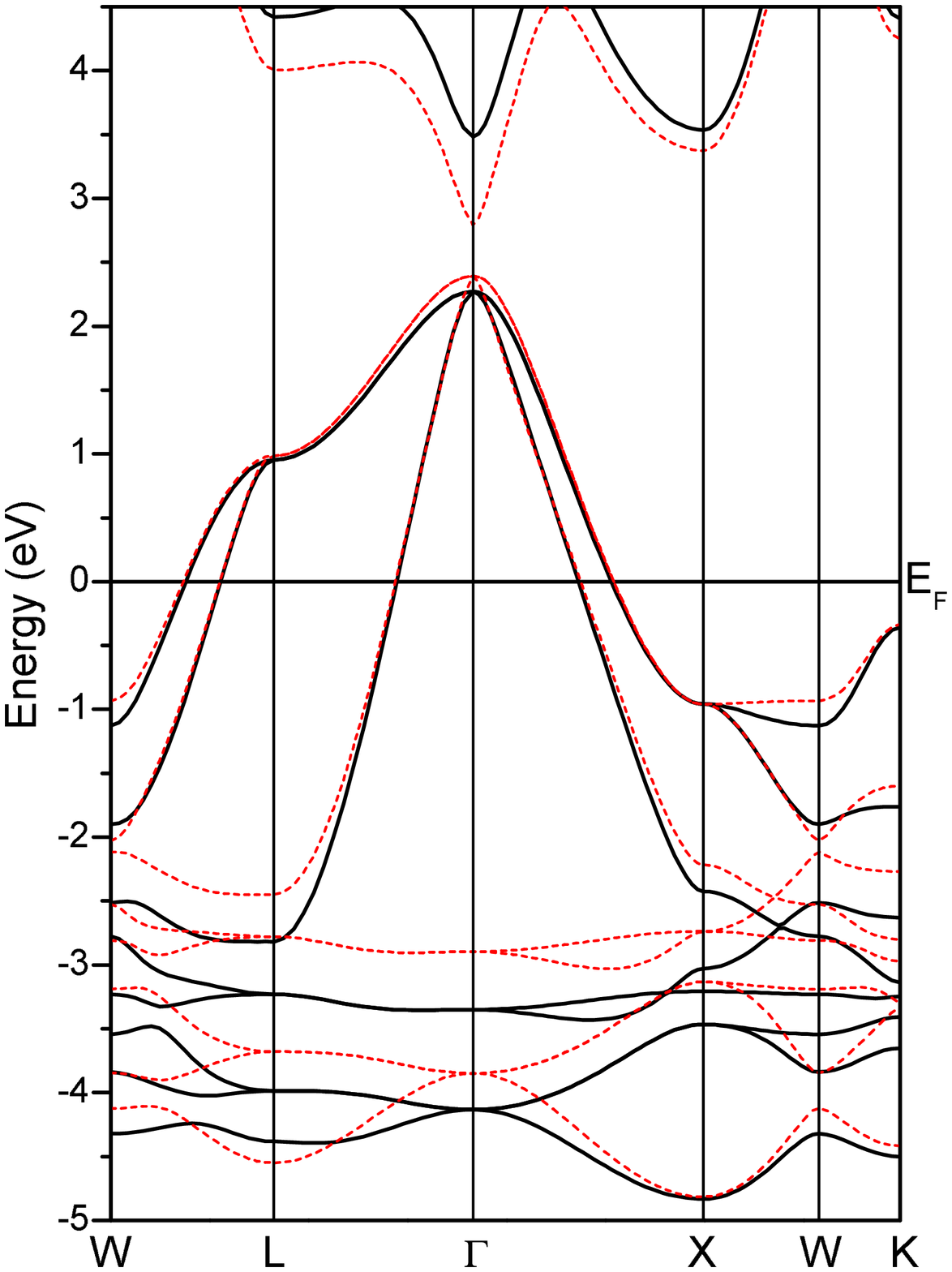}
\includegraphics[width=4.2cm]{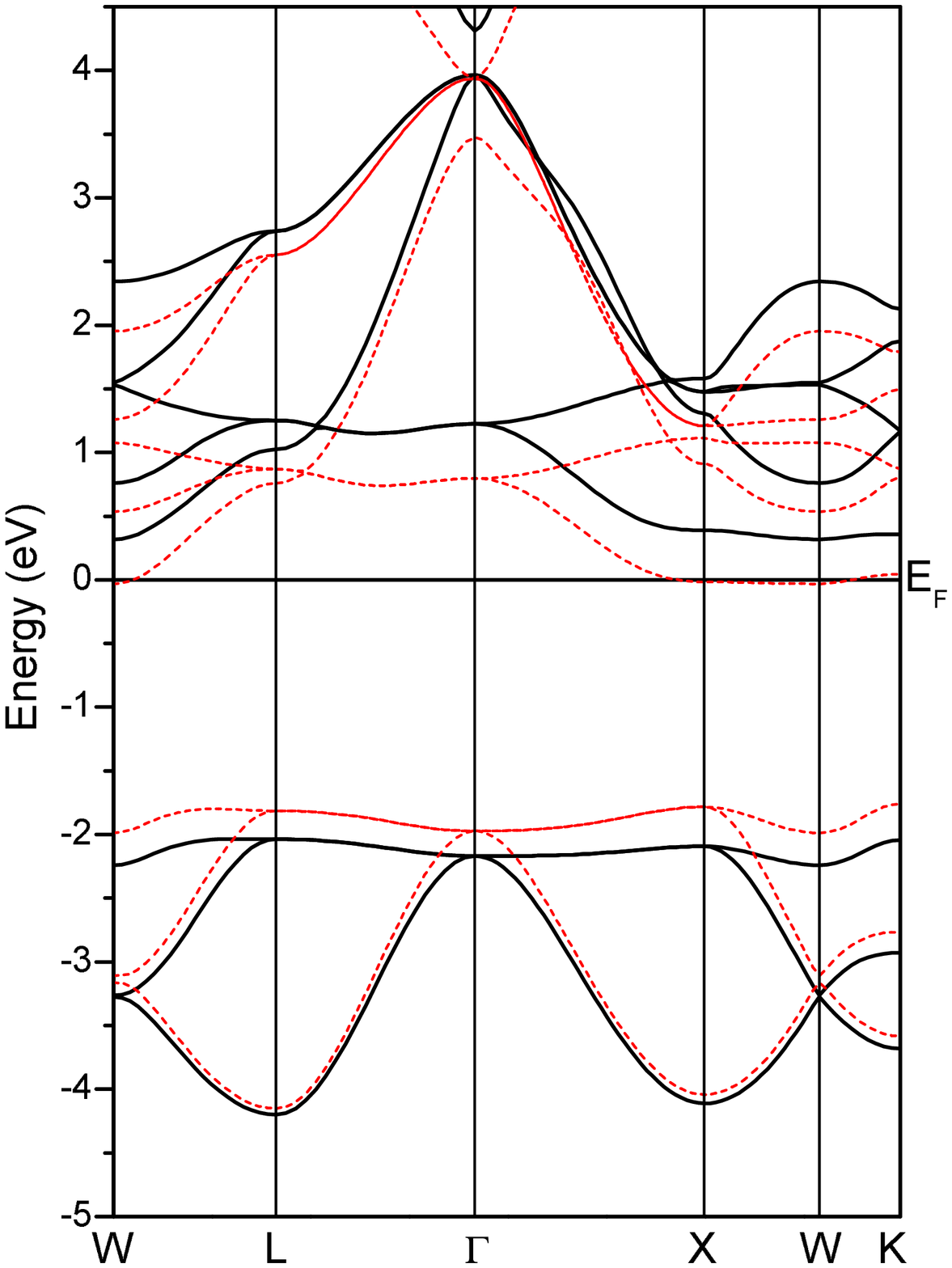}
\caption{(color online) Spin-resolved energy bands of MnAs with
the zincblende structure, calculated with mBJLDA (solid lines) and
GGA-PBE (dashed lines), respectively. The left panel is for
majority-spin channel and the right panel for
minority-spin.}\label{mnas-eb}
\end{center}
\end{figure}

\begin{figure}[!htb]
\begin{center}
\includegraphics[width=4.3cm]{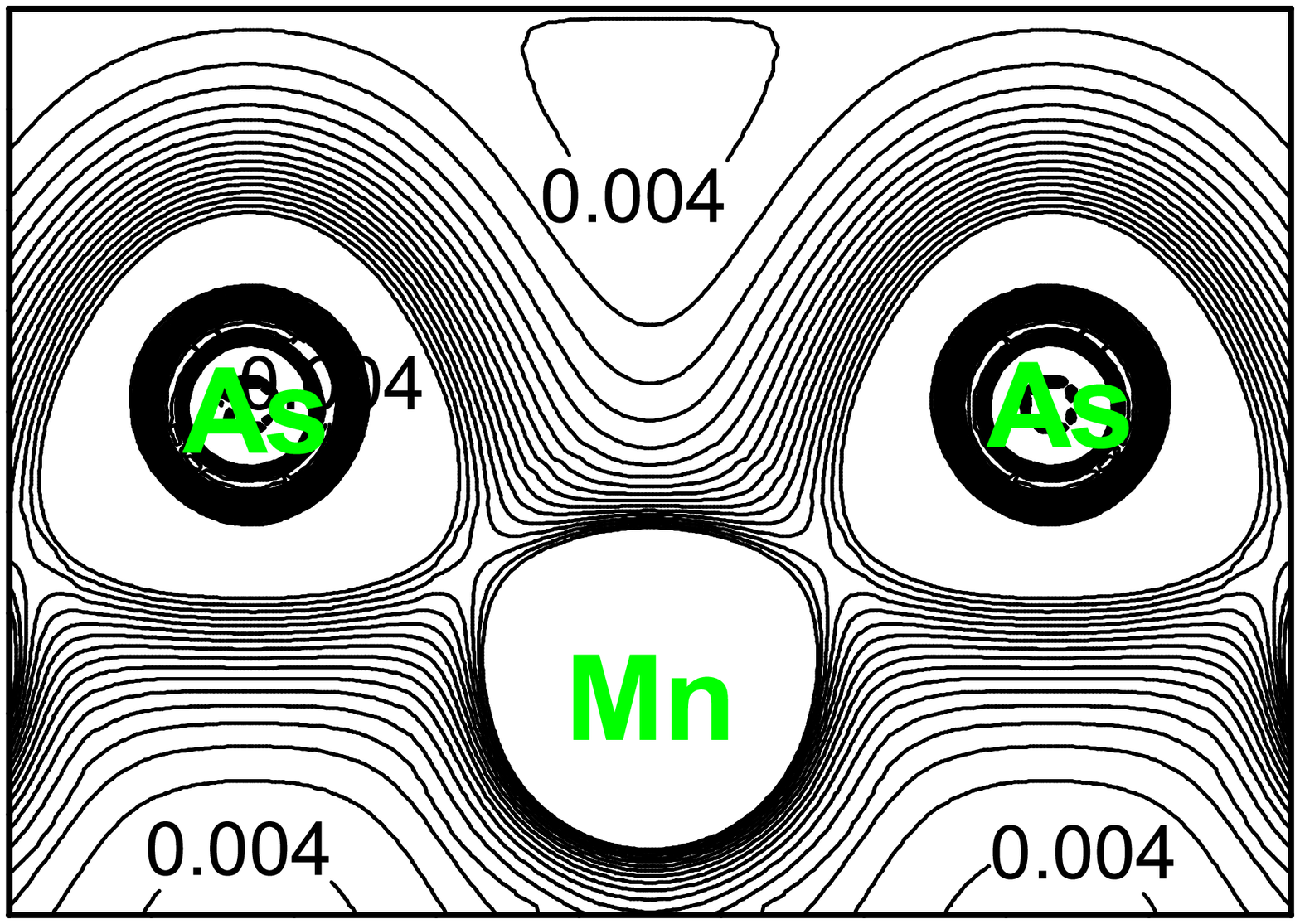}
\includegraphics[width=4.3cm]{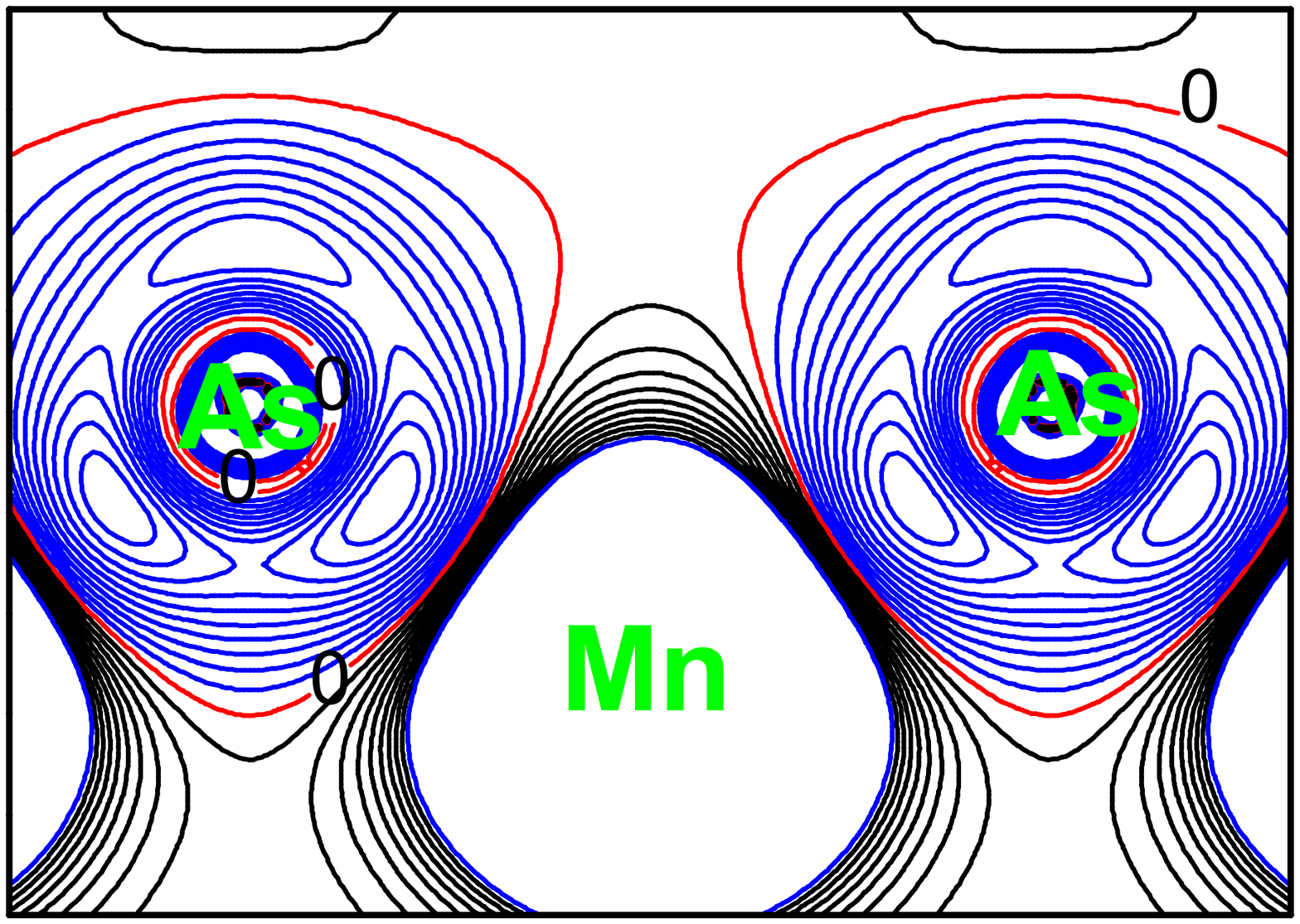}
\caption{(color online) The charge (upper panel) and magnetization
(lower panel) density distributions of MnAs with the zincblende
structure. The plane is defined by an As-Mn-As triangle. Upper:
The actual charge density is 0.004$\sim$0.068 $|e|$/bohr$^3$ and
the density increment is 0.004$|e|$/bohr$^3$. Lower: The lines
with '0' label the zero value for the magnetization density
(-0.008$\sim$+0.008 $|\mu_B|$/bohr$^3$) and the increment is
0.0008$\mu_B$/bohr$^3$.}\label{ele}
\end{center}
\end{figure}

In Fig. 2 we present the spin-resolved energy bands between -5 and
4.5 eV of zincblende MnAs calculated with mBJLDA and GGA-PBE.
There are little differences between mBJLDA and GGA for the
majority-spin $t_{2g}$ bands near the Fermi level, but the mBJLFA
raises the minority-spin $e_g$ bands by 0.3 eV and lowers the
majority-spin $e_g$ bands by 0.3 eV. These are vital to make
zincblende MnAs become a truly half-metallic ferromagnet. We
present the charge and magnetization density distributions of the
zincblende MnAs calculated with the mBJLDA potential, within the
charge density range 0 $\sim$ 0.068 $|e|$/bohr$^3$ and the
magnetization density -0.008 $\sim$ +0.008 $|\mu_B|$/bohr$^3$,
respectively. Here the energy range is from -5 eV to the Fermi
level and the plane is defined by the nearest As-Mn-As triangle.
There is not any visible difference if mBJLDA is replaced by GGA.
It can be seen that the Mn-As bond is covalent and there is a
small negative magnetization around As although the total magnetic
moment in the cell is contributed mainly by Mn.

\begin{figure*}[!htb]
\begin{center}
\includegraphics[width=5.6cm]{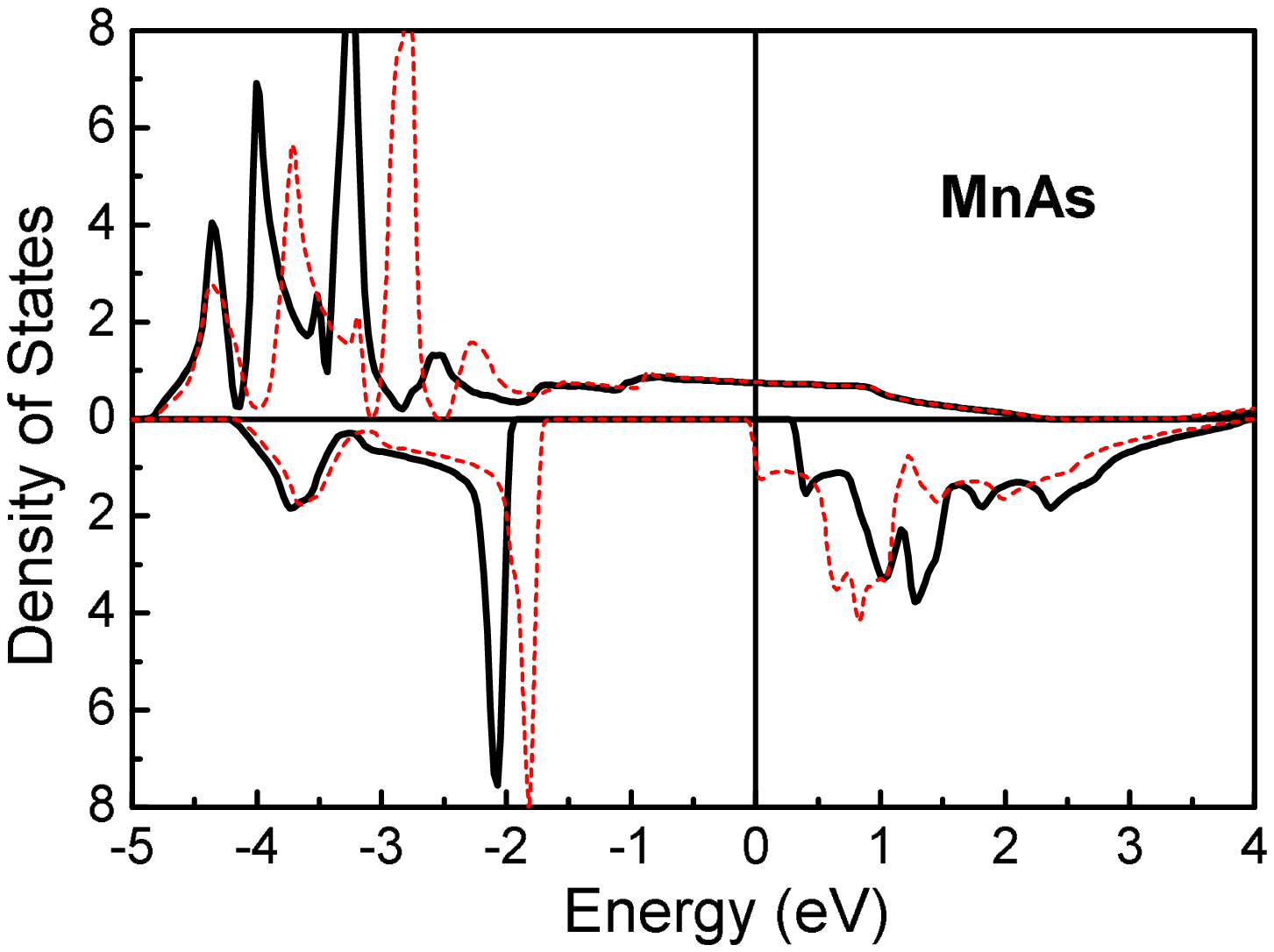}
\includegraphics[width=5.6cm]{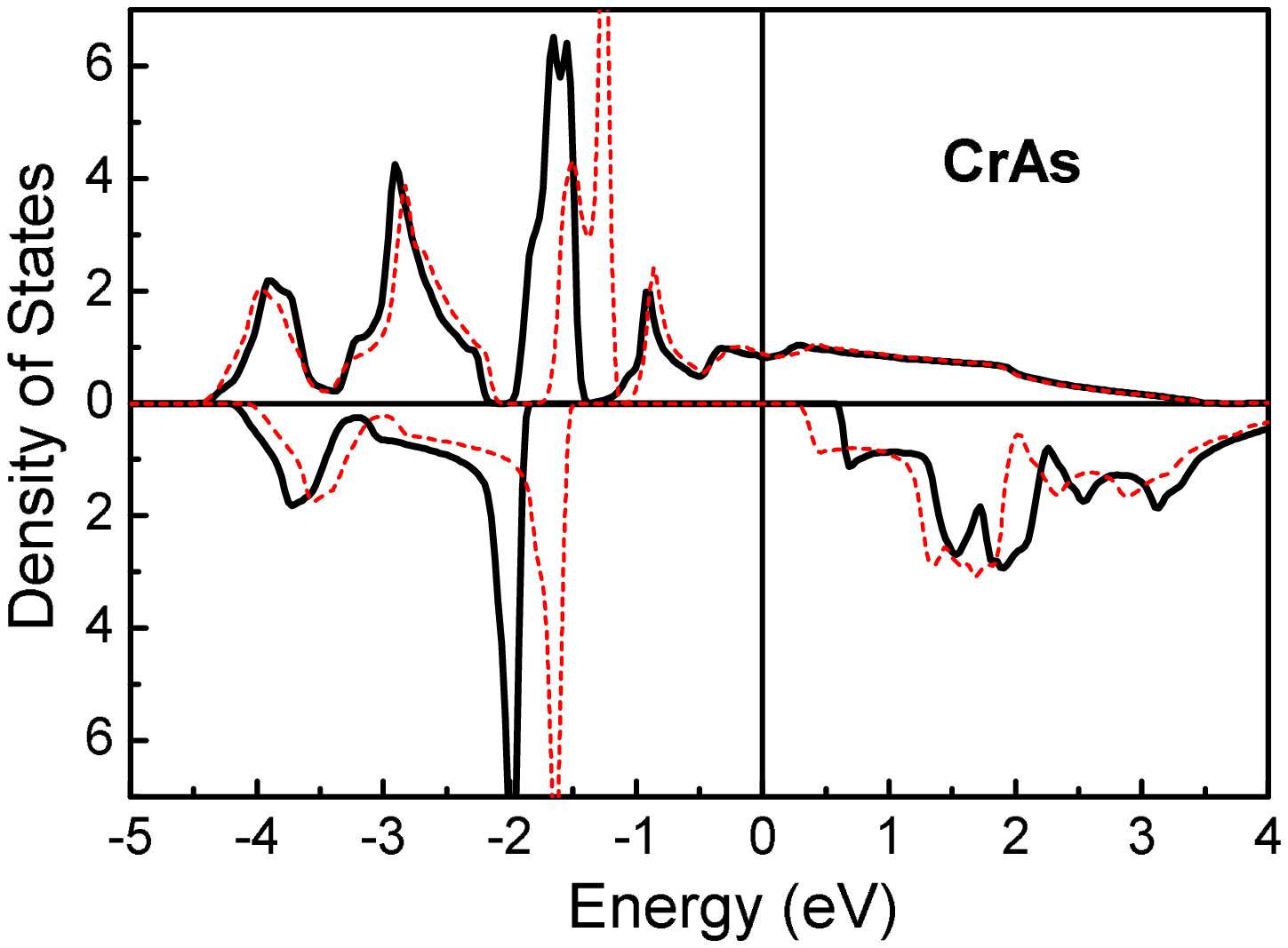}
\includegraphics[width=5.6cm]{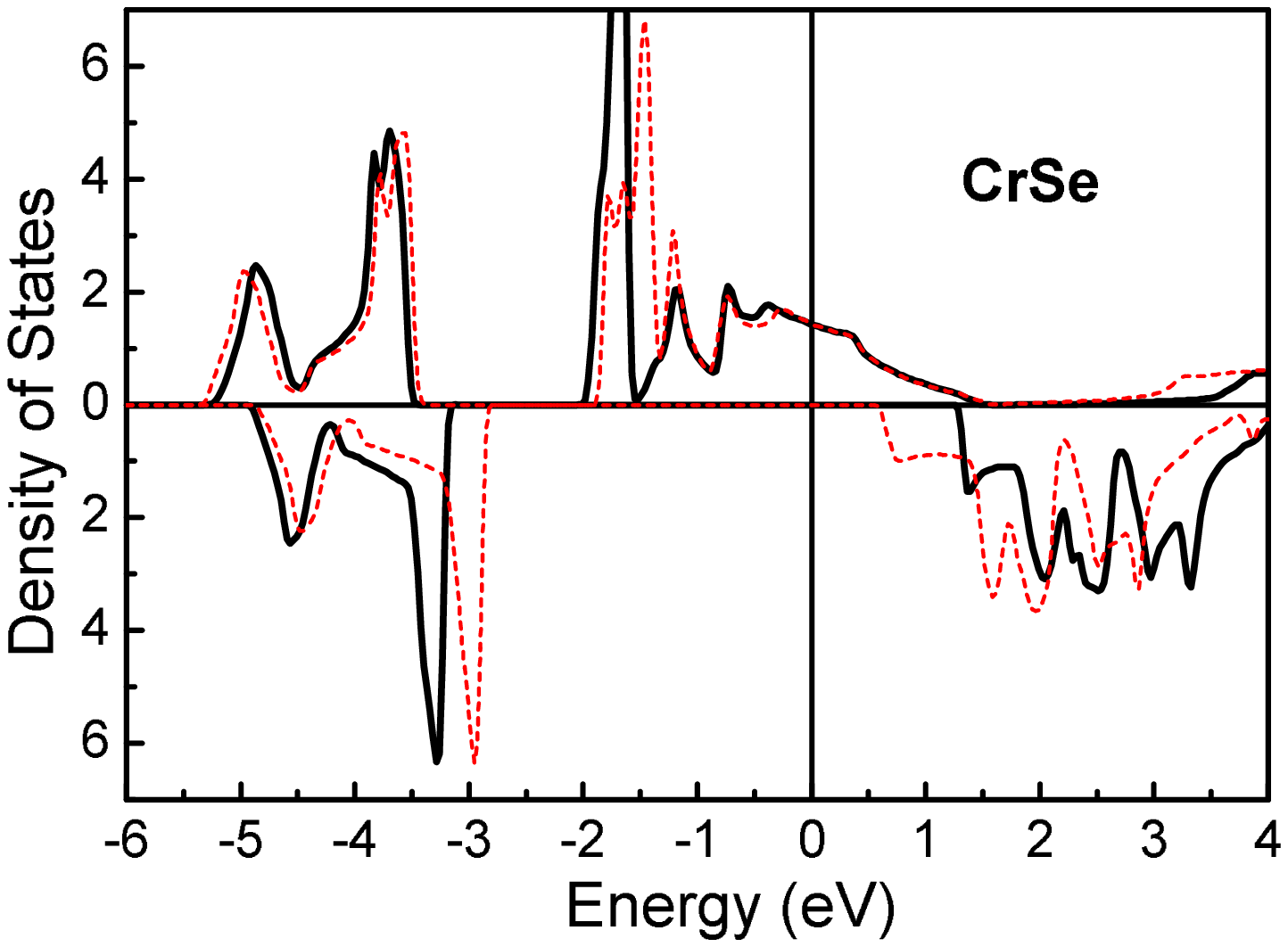}\\
\includegraphics[width=5.6cm]{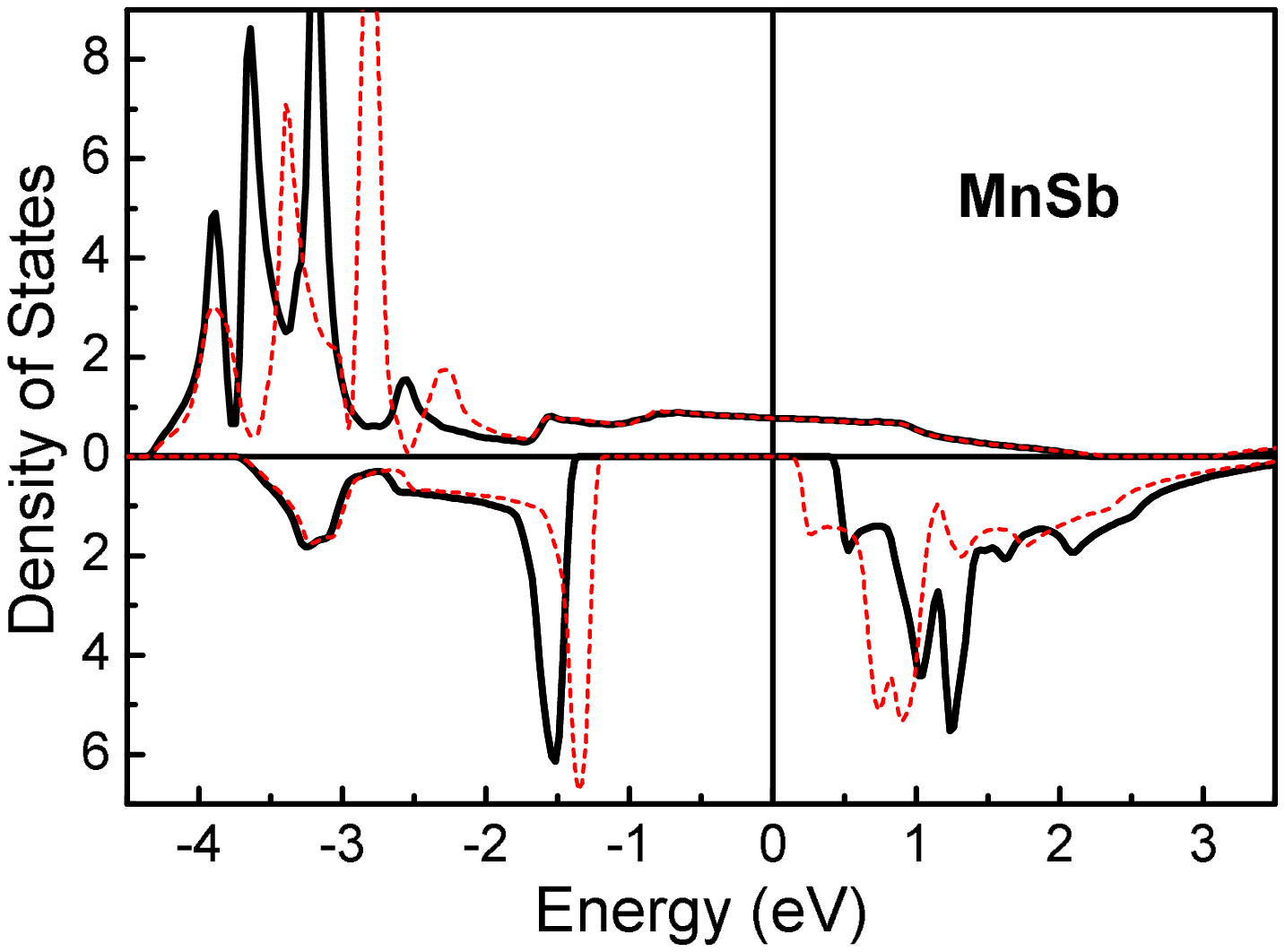}
\includegraphics[width=5.6cm]{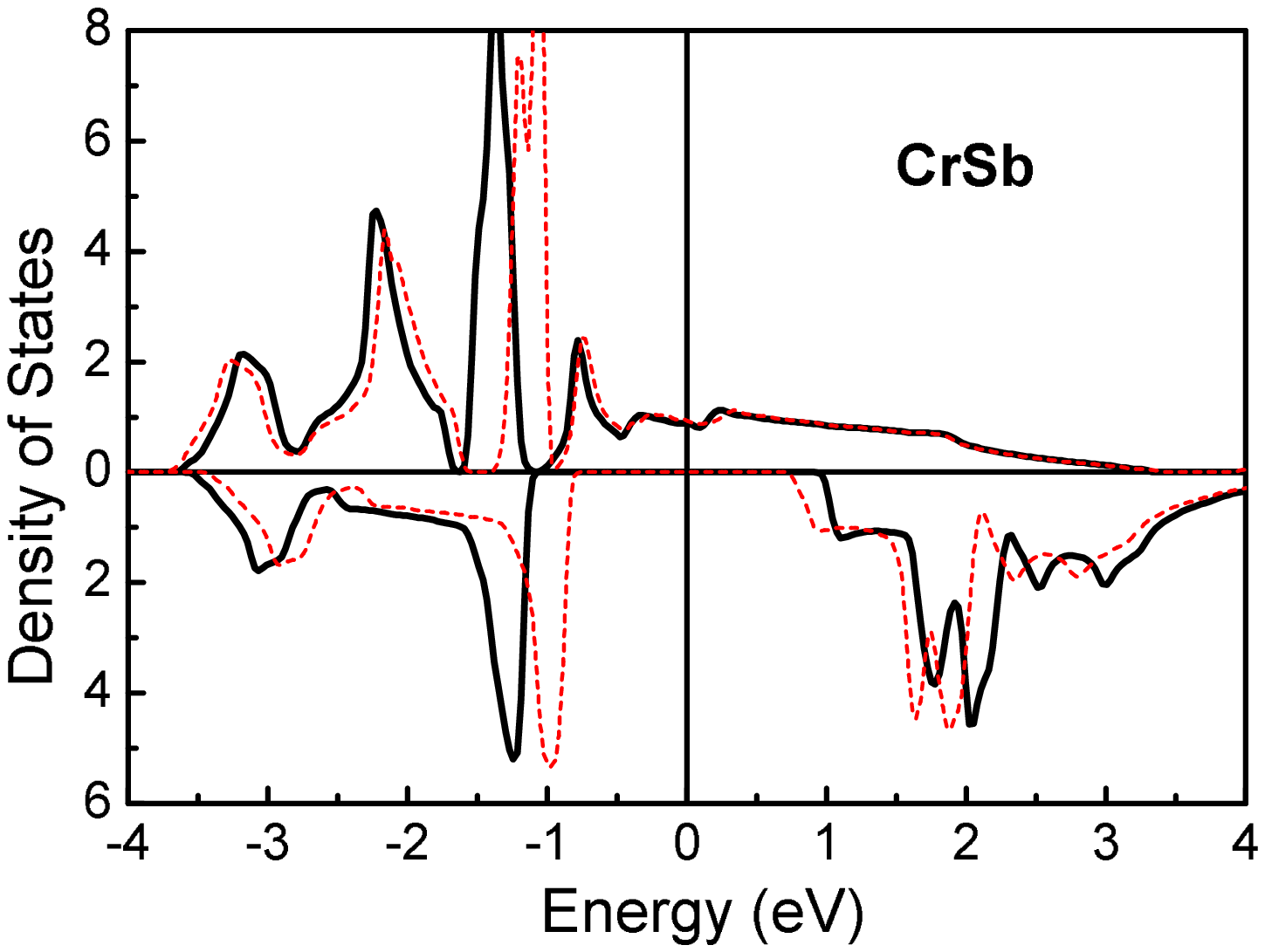}
\includegraphics[width=5.6cm]{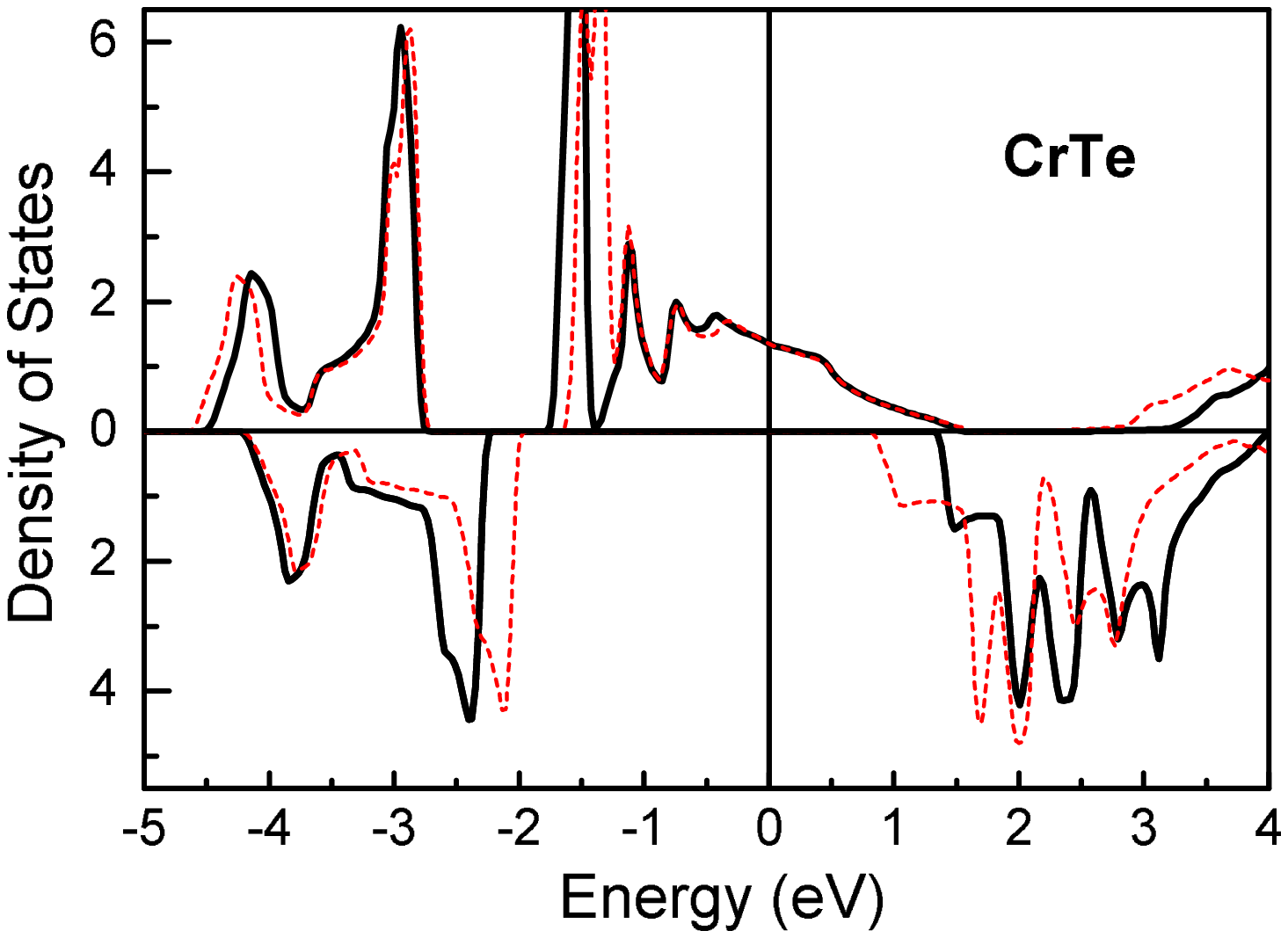}
\caption{(color online) Spin-resolved density of states of MnAs
(left-upper), MnSb (left-lower), CrAs (middle-upper), CrSb
(middle-lower), CrSe (right-upper), and CrTe (right-lower) with
the zincblende structure, calculated with mBJLDA (solid lines) and
GGA-PBE (dashed lines), respectively. The upper part in each panel
is for majority-spin channel and the lower part for
minority-spin.}\label{dos} \end{center}
\end{figure*}

The mBJLDA potential has similar effect on the electronic
structures of other zincblende compounds, such as MnSb, CrAs,
CrSb, CrSe, and CrTe. We present the spin-resolved total DOSs of
zincblende MnAs, MnSb, CrAs, CrSb, CrSe, and CrTe calculated with
GGA-PBE and mBJLDA. For convenience of comparison, we plot the
GGA-PBE and mBJLDA DOSs in the same panel for each of the six
zincblende compounds. It is clear that in each of the six cases
the minority-spin gap across the Fermi level is made larger and
the majority-spin $e_g$ bands move downward when GGA-PBE is
replaced by mBJLDA. Furthermore, the bottom of the minority-spin
conduction bands calculated with mBJLDA is higher than that with
GGA, and consequently, the half-metallic gap is effectively made
larger by replacing GGA with mBJLDA. All the mBJLDA results of the
magnetic moments ($M$), the minority-spin gaps ($G_{\rm mis}$),
and the half-metallic gaps ($G_{\rm hm}$) are summarized in Table
\ref{t1}. For comparison, we also present the corresponding
GGA-PBE results in the parentheses. It is clear that both $G_{\rm
mis}$ and $G_{\rm hm}$ are made larger by replacing GGA-PBE with
mBJLDA. Most importantly, zincblende MnAs is a truly half-metallic
ferromagnet if mBJLDA is used. Therefore, the half-metallic
ferromagnetism has been improved after we use mBJLDA instead of
GGA-PBE.

\begin{table}[htb]
\caption{The lattice constants ($a$), magnetic moments per formula
unit ($M$), minority-spin gaps ($G_{\rm mis}$), and half-metallic
gaps ($G_{\rm hm}$) of zincblende MnAs, MnSb, CrAs, CrSb, CrSe,
and CrTe calculated with mBJLDA. The values in the parentheses are
calculated with GGA-PBE.}
\begin{tabular*}{0.48\textwidth}{@{\extracolsep{\fill}}ccccc}
\hline\hline Name  & $a$ (\AA) & $M$ ($\mu_B$) & $G_{\rm mis}$ (eV) & $G_{\rm hm}$ (eV)\\
\hline MnAs & 5.717 &  4 (3.8) &  2.352 (1.760) & 0.318 ( - )\\
\hline MnSb & 6.166 &  4 (4) &  1.923 (1.491) & 0.459 (0.200)\\
\hline CrAs & 5.659 &  3 (3) &  2.579 (1.896) & 0.636 (0.322)\\
\hline CrSb & 6.138 &  3 (3) &  2.180 (1.650) & 0.992 (0.751)\\
\hline CrSe & 5.833 &  4 (4) &  4.569 (3.508) & 1.331 (0.604)\\
\hline CrTe & 6.292 &  4 (4) &  3.692 (2.919) & 1.384 (0.858)\\
\hline\hline
\end{tabular*}\label{t1}
\end{table}

\section{Discussions}

It has been proved that mBJLDA potential improves semiconductor
gaps for various semiconducotrs\cite{mbj09}. Our calculated
results have shown that the minority-spin gap across the Fermi
level and the half-metallic gap (the key parameter\cite{lbgb}) of
the binary transition-metal pnictides and chalcogenides are both
improved by mBJLDA potential. Especially, the nearly-half-metallic
zincblende MnAs under GGA and LDA becomes a truly half-metallic
ferromaget when we use mBJLDA potential. This is consistent with
experimental result that zincblende MnAs is a truly half-metallic
ferromagnet\cite{mnas,mnas1,mnas2}. Therefore, we believe that
mBJLDA potential as an exchange-correlation potential is more
satisfactory than both GGA and LDA in describing half-metallic
ferromagents akin to corresponding semiconductors, such as
zincblende MnAs, MnSb, CrAs, CrSb, CrSe, and CrTe. More
fortunately, the computational efficiency of mBJLDA is almost the
same as those of GGA and LDA, and hence one can calculate large
systems with mBJLDA as one does with GGA and LDA.

What causes this improvement? On one hand, our calculated results
show that the replacing GGA with mBJLDA has little effect on the
wide majority-spin bands near the Fermi level. Actually, mBJLDA
does not yield visible changes in wide transition-metal d-$t_{2g}$
bands with small density of states, but indeed shifts the narrow
$e_g$ bands with respect to the $t_{2g}$ bands. We can see in Fig.
4 that the occupied majority-spin $e_g$ bands emerge with the
lower occupied majority-spin p-bands in the cases of MnAs and
MnSb. On the other hand, the occupied minority-spin p-bands are
lowered by 0.25$\sim$0.35 eV and the empty (or nearly empty)
minority-spin $e_g$ bands are raised by 0.33$\sim$0.73 eV, which
enhances substantially both the majority-spin gap and the
half-metallic gap. These can be understood by assuming it is the
mBJLDA potential that enhances the spin exchange splitting with
respect to GGA and LDA results. This explanation is supported by
the observation in Fig. 1 that there are much d-$t_{2g}$ weight in
the top of the minority-spin valence bands and in the bottom of
the minority-spin conduction bands. The enhancement of the spin
exchange splitting is achieved because of the advantage of the BJ
exchange potential\cite{bj,mbj09}. Thus, the half-metallic
ferromagnetism is enhanced in the binary transition-metal
pnictides and chalcogenides akin to semiconductors.

\section{Conclusion}

In summary, we have used a state-of-the-art DFT method with the
mBJLDA potential\cite{mbj09,pw92} to investigate the electronic
structures of zincblende transition-metal pnictides and
chalcogenides and compare the mBJLDA results with those calculated
with popular exchange-correlation potentials (GGA-PBE and
LDA-PW91). Our calculated results show that the mBJLDA potential
does not yield visible changes in wide transition-metal d-$t_{2g}$
bands near the Fermi level, but makes the occupied minority-spin
p-bands lower by 0.25$\sim$0.35 eV and the empty (or nearly empty)
minority-spin $e_g$ bands higher by 0.33$\sim$0.73 eV. In the
cases of zincblende MnSb, CrAs, CrSb, CrSe, and CrTe, the
half-metallic gaps are enhanced by 19$\sim$56\% with respect to
the GGA-PBE and LDA-PW91 results. Most importantly, the mBJLDA
potential makes zincblende MnAs become a truly half-metallic
ferromagnet. This result is consistent with experiment, which is a
direct evidence that the mBJLDA potential is more satisfactory
than popular GGA and LDA in describing the electronic structures
of the binary transition-metal pnictides and chalcogenides akin to
semiconductors. The improved half-metallic ferromagnetism can be
naturally understood in terms of the enhanced spin exchange
splitting caused by the mBJLDA potential. We believe that the
semi-local and orbital-independent mBJLDA potential without
atom-dependent parameters will be more satisfactory than usual GGA
and LDA in calculating electronic structures with almost the
computational efficiency of usual GGA and LDA for magnetic
materials sharing some features with semiconductors.

\acknowledgments This work is supported  by Nature Science
Foundation of China (Grant Nos. 10874232 and 10774180), by the
Chinese Academy of Sciences (Grant No. KJCX2.YW.W09-5), and by
Chinese Department of Science and Technology (Grant No.
2005CB623602).

\end{document}